\documentclass{PoS}

\title{Kaon $B$-parameters for Generic $\Delta S=2$
Four-Quark Operators in Quenched Domain Wall QCD}
\ShortTitle{Kaon $B$-parameters for Generic $\Delta S=2$
Four-Quark Operators}

\author{CP-PACS Collaboration:
 \speaker{Y.~Nakamura}${}^{a}$\thanks{E-mail: nakayou@het.ph.tsukuba.ac.jp}~,
 S.~Aoki${}^{a,b}$,
 M.~Fukugita${}^{c}$, 
 N.~Ishizuka${}^{a,e}$,
 Y.~Iwasaki${}^{a}$,
 K.~Kanaya${}^{a}$,
 Y.~Kuramashi${}^{a,e}$,
 J.~Noaki${}^{f}$,
 M.~Okawa${}^{d}$,
 Y.~Taniguchi${}^{a,e}$,
 A.~Ukawa${}^{a,e}$,
 T.~Yoshi\'e${}^{a,e}$,
 \\
 \llap{${}^a$}Graduate School of Pure and Applied Sciences, University of Tsukuba, Tsukuba, Ibaraki 305-8571, Japan\\
 \llap{${}^b$}Riken BNL Research Center, Brookhaven National Laboratory, Upton, New York 11973, USA\\
 \llap{${}^c$}Institute for Cosmic Ray Research, University of Tokyo,  Kashiwa, Chiba 277-8572, Japan\\
 \llap{${}^d$}Department of Physics, Hiroshima University, Higashi-Hiroshima, Hiroshima 739-8526, Japan\\
 \llap{${}^e$}Center for Computational Sciences, University of Tsukuba, Tsukuba, Ibaraki 305-8577, Japan\\
 \llap{${}^f$}School of Physics and Astronomy, University of Southampton, Southampton, SO17 1BJ, UK}

\abstract{
We present a study of $B$-parameters for generic $\Delta S=2$
 four-quark operators in domain wall QCD.
Our calculation covers all the $B$-parameters required
to study the neutral kaon mixing in the standard model (SM) and beyond it.
We evaluate one-loop renormalization factors of 
the operators employing the plaquette and Iwasaki 
gauge actions. Numerical simulations are carried out in quenched QCD
with both gauge actions on $16^3\times 32\times 16$ and 
$24^3\times 32\times 16$ at the lattice spacing
$1/a\approx 2$GeV.
We investigate the relative magnitudes of the
non-SM $B$-parameters to the SM one, which are compared with
the previous results obtained with the overlap and the clover quark actions.
}

\FullConference{XXIVth International Symposium on Lattice Field Theory\\
		 July 23-28, 2006\\
		 Tucson, Arizona, USA}

\begin{document}

\section{Introduction}
%
The decay processes of the neutral kaon system
contain rich physics in the SM and beyond it.
The indirect CP violation parameter $\epsilon$
provides us a chance to examine the SM and make constraints 
on the physics beyond it.
In order to determine $\epsilon$ we need the kaon matrix element
of the effective Hamiltonian for $\Delta S=2$ transition, whose
general structure is written as
$\langle {\bar K}^0|H_{\rm eff}^{\Delta S=2}|K^0\rangle$,
where
$H_{\rm eff}^{\Delta S=2}={G_F}/{\sqrt{2}}
\sum_i V^i_{\rm CKM} C_i(\mu){\cal O}_i$
with $G_F$ the Fermi constant, $V^i_{\rm CKM}$ the 
CKM factors, $C_i$ the Wilson coefficients  
and ${\cal O}_i$ the relevant local operators for the process.
Although it is known that only the left-left operator ${\cal O}_{\rm LL}$ 
is relevant in the SM, we need the operators 
with more general chiral structures for the physics beyond the SM.
For example  the supersymmetric model requires\cite{fcnc_susy}
\begin{eqnarray}
{\cal O}_1 &=&
\bar{s}^a\gamma_\mu(1-\gamma_5)d^a\bar{s}^b\gamma_\mu(1-\gamma_5)d^b,
\label{eq:op_susy1} \\
{\cal O}_2 &=&
\bar{s}^a(1-\gamma_5)d^a\bar{s}^b(1-\gamma_5)d^b,  
\label{eq:op_susy2}\\
{\cal O}_3 &=&
\bar{s}^a(1-\gamma_5)d^b\bar{s}^b(1-\gamma_5)d^a, 
\label{eq:op_susy3}\\
{\cal O}_4 &=&
\bar{s}^a(1-\gamma_5)d^a\bar{s}^b(1+\gamma_5)d^b,
\label{eq:op_susy4}\\
{\cal O}_5 &=&
\bar{s}^a(1-\gamma_5)d^b\bar{s}^b(1+\gamma_5)d^a
\label{eq:op_susy5}
\end{eqnarray}
and $\tilde{\cal O}_{1,2,3}$ obtained 
from the ${\cal O}_{1,2,3}$ by exchanging the
left- and right-handed quarks.
${\cal O}_1$ is the same operator in the SM,
while ${\cal O}_{2,3,4,5}$ and $\tilde{\cal O}_{1,2,3}$
are the non-SM operators.
 
In this report we present the results of 
$\langle {\bar K}^0|{\cal O}_i|K^0\rangle$
($i=1,\dots,5$) obtained in quenched domain wall QCD.
Although these operators are introduced by the supersymmetric
model, we should note that 
$\langle {\bar K}^0|{\cal O}_i|K^0\rangle$ with $i=1,\dots,5$ 
are a complete set of the kaon matrix elements of the four-quark operators
allowed by the symmetries.
This implies that they are sufficient for analysis of any model beyond the SM.
We are particularly interested in the relative magnitude of non-SM to 
SM matrix elements:
\begin{eqnarray}
R_i(\mu) \equiv \frac{\langle{\bar K}^0|{\cal O}_i(\mu)|K^0\rangle}
 {\langle{\bar K}^0|{\cal O}_1(\mu)|K^0\rangle},
\hspace{0.5cm} i=2,\dots,5.
\label{eq:ratio}
\end{eqnarray}
We also present the kaon $B$-parameter for the SM:
\begin{eqnarray}
B_1(\mu) &=& B_K(\mu) \equiv
\frac{\langle {\bar K}^0|{\cal O}_1(\mu)|K^0\rangle}
{\frac{8}{3}\langle
{\bar K}^0|\bar{s}\gamma_\mu\gamma_5d|0\rangle\langle0|\bar{s}\gamma_\mu\gamma_5d|K^0\rangle},
\label{eq:Bk}
\end{eqnarray}
and those for the non-SM:
\begin{eqnarray}
B_i(\mu) \equiv
 \frac{\langle{\bar K}^0|{\cal O}_i(\mu)|K^0\rangle}
{C_i\langle{\bar K}^0|\bar{s}\gamma_5d(\mu)|0\rangle 
\langle 0|\bar{s}\gamma_5d(\mu)|K^0\rangle}
,\hspace{0.5cm} i=2,\dots,5, 
\label{eq:Bi}
\end{eqnarray}
where $C_i\equiv \{\frac{5}{3},-\frac{1}{3},-2,-\frac{2}{3}\}$ 
are the convention factors introduced by the vacuum saturation
approximation.
Our results are compared with the previous works employing the
overlap\cite{Babich:2006bh} and the clover quark actions\cite{Allton:1998sm}.

\section{Simulation details}
\subsection{Simulation parameters}

In Table~\ref{tab:param} we summarize the simulation parameters, part of
which are the same as in the previous CP-PACS
calculation of $\epsilon'/\epsilon$\cite{Noaki:2001un}.
The domain wall fermion with the wall height $M=1.8$
is employed with the plaquette and the Iwasaki gauge actions 
in quenched approximation.
We choose $N_s\times N_t\times N_5=16^3\times32\times16$ 
lattice at $\beta=6.0$ for the plaquette
gauge action, while $N_s=16^3$ and $24^3$ 
at $\beta=2.6$ for the Iwasaki gauge action 
in order to investigate the volume dependence.
As for the lattice spacing we refer to Refs.\cite{necco,takeda}
where the lattice spacing determined from the Sommer scale $r_0=0.5$fm
is parameterized in terms of $\beta$.
We take degenerate masses $m_{f}a=0.02,0.03,0.04,0.05,0.06$ 
for the up, down and strange quarks in both gauge actions.
For each value of $m_f a$ gauge configurations are independently 
generated with the algorithm of the five-hit pseudo-heat-bath combined with
four overrelaxation sweeps, which we call an iteration.
In Table~\ref{tab:param} we list the numbers of configurations with the
interval of 200 iterations for measurements.
The anomalous quark masses $m_{5q}$ for both gauge actions are
calculated in Ref.\cite{dwf_chiral}.
Their values in the physical unit given in Table~\ref{tab:param} 
are obtained by using the lattice spacing discussed above.

\begin{table}[t]
\begin{center}
\caption{Simulation parameters. $P$ and $R$ denote the expectation
 values for the plaquette and the $1\times 2$ rectangular.}
\label{tab:param}
\begin{tabular}{c|c|cc}
\hline
\hline
& plaquette & \multicolumn{2}{c}{Iwasaki}  \\
\hline
$\beta$ & 6.0 & \multicolumn{2}{c}{2.6}  \\
$N_s\times N_t\times N_5$ & $16^3\times32\times16$    & $16^3\times32\times16$ & $24^3\times32\times16$  \\ 
$a^{-1}$[GeV] & 2.12\cite{necco} & \multicolumn{2}{c}{2.00\cite{takeda}}  \\
$P$           & 0.59374\cite{pt_dwf}   & \multicolumn{2}{c}{0.670632(10)\cite{dwf_bk}} \\
$R$           &  $-$      & \multicolumn{2}{c}{0.45283(2)\cite{dwf_bk}} \\
No. config. & 180  & 400  & 200 \\ 
   & & &(400 for $m_f a=0.02$)\\
$m_{5q}$[MeV] & 3.14(36) & 0.292(44)  & $-$ \\
\hline
\hline
\end{tabular}
\end{center}
\end{table}

\subsection{Calculational method}

The ratios $R_i$ are extracted from the plateau of
\begin{eqnarray}
\frac{\sum_{{\vec z},{\vec y},{\vec x}}\langle{\bar K}^0({\vec z},N_t-1) {\cal O}_i({\vec y},t) K^0({\vec x},0)\rangle}
{\sum_{{\vec x},{\vec y},{\vec z}}\langle{\bar K}^0({\vec z},N_t-1) {\cal O}_1({\vec y},t) K^0({\vec x},0)\rangle},
\end{eqnarray}
choosing $0\ll t\ll N_t-1$.
Similarly $B_K$ and 
$B_i$ ($i=2,\cdots,5$) are obtained from the three-point functions
divided by two-point functions,
\begin{eqnarray}
\frac{\sum_{{\vec z},{\vec y},{\vec x}}
\langle{\bar K}^0({\vec z},N_t-1) {\cal O}_i({\vec y},t) K^0({\vec x},0)\rangle}
{\frac{8}{3}\sum_{{\vec z},{\vec y}}\langle{\bar K}^0({\vec z},N_t-1) A_0({\vec y},t)\rangle 
\sum_{{\vec y}^\prime,{\vec x}}\langle A_0({\vec y}^\prime,t) K^0({\vec x},0)\rangle}
&\hspace{3mm}{\rm for}\hspace{2mm} & B_K,
\\
\frac{\sum_{{\vec z},{\vec y},{\vec x}}
\langle{\bar K}^0({\vec z},N_t-1) {\cal O}_i({\vec y},t) K^0({\vec x},0)\rangle}
{C_i\sum_{{\vec z},{\vec y}}\langle{\bar K}^0({\vec z},N_t-1) P({\vec y},t)\rangle 
\sum_{{\vec y}^\prime,{\vec x}}\langle P({\vec y}^\prime,t) K^0({\vec x},0)\rangle}
&\hspace{3mm}{\rm for}\hspace{2mm} & B_i.
\end{eqnarray}
In order to calculate the hadron correlation functions we
solve quark propagators with the Coulomb gauge fixing employing wall
sources placed at the edges of lattice where the Dirichlet boundary
condition is imposed in the time direction.

We estimate errors by the single elimination jackknife procedure for all
measured quantities.

\subsection{Renormalization factors}
In order to convert the matrix elements obtained on the lattice 
to those defined in the continuum,
we make a one-loop perturbative matching of the lattice operators
and the continuum ones at a scale $\mu=1/a$, where the latter is
defined in the $\overline{\rm MS}$ scheme with the naive dimensional
regularization(NDR).
Since the domain wall quarks retains the good chiral
symmetry on the lattice, the four-quark operators are renormalized 
in the same way as in the continuum:
multiplicative renormalization for ${\cal O}_1$, while 
mixing between ${\cal O}_2$ and ${\cal O}_3$ and 
between ${\cal O}_4$ and ${\cal O}_5$.

The one-loop perturbative renormalization factors with 
mean field improvement are written as\cite{pt_dwf}
\begin{eqnarray}
{\cal O}_i^{\overline{\rm MS}}(\mu)
=\frac{1}{(1-w_0)^2Z_w^2}Z_{ij}(\mu a){\cal O}_j(1/a)^{\rm lat},
\end{eqnarray}
where $w_0$ is a function of domain wall height, 
$Z_w$ is the renormalization factor of domain wall height 
and $Z_{ij}$ are those of the four-quark operators.
Indices $i,j$ label the operator number.
Numerical values of $Z_{ij}$ with mean field
improvement for the Iwasaki and the plaquette gauge actions
are given in Ref.\cite{Nakamura:2006zx}.
Employing $\beta=6.0$, $a^{-1}=2.12$GeV and $P=0.59374$ 
for the plaquette gauge action we replace 
$M\rightarrow {\tilde M}=M+4(u-1)=1.311$ with $u=P^{1/4}$
and obtain
\begin{eqnarray}
           Z_{ij}^{\rm plaquette}(\mu=1/a)&=&
        \left(
         \begin{array}{@{\,}ccccc@{\,}}
           0.7287 &0&0&0&0 \\
           0& 0.6845& -0.00156&0&0 \\
           0& 0.0352& 0.8682 &0&0 \\
           0&0&0& 0.6325& -0.0414 \\
           0&0&0& -0.0689& 0.7564  \\
         \end{array}
         \right).
\end{eqnarray}
For the Iwasaki gauge action we obtain $\tilde{M}=1.420$ and
\begin{eqnarray}
           Z_{ij}^{\rm Iwasaki}(\mu=1/a)&=&
        \left(
         \begin{array}{@{\,}ccccc@{\,}}
           0.8062 &0&0&0&0 \\
           0& 0.8124& -0.00679 &0&0 \\
           0& 0.0156 &0.9241 &0&0 \\
           0&0&0&0.7847&-0.0427 \\
           0&0&0&-0.0477&0.8425 \\
         \end{array}
         \right).
\end{eqnarray}
for $\beta=2.6$, $a^{-1}=2.00$GeV, $P=0.670632(10)$ and $R=0.45283(2)$.
The details of the mean field improvement for the domain wall fermion
are explained in Refs.\cite{pt_dwf,Nakamura:2006zx}.
We observe that the Iwasaki gauge action shows smaller one-loop
corrections than the plaquette gauge action.
For the axial vector current and the pseudoscalar density
in $B_K$ and $B_i$ ($i=2,\dots,5$), we also use the one-loop
perturbative renormalization factors with mean field improvement.


\section{Results for $R_i$ and $B_i$}

Since the kaon matrix element of ${\cal O}_1$ is proportional to $M_K^2$,
the ratios $R_i$ in Eq.(\ref{eq:ratio}) should diverge toward the chiral limit.
In order to keep the ratios finite in the chiral limit and 
tame the quark mass dependences of $R_i$, we introduce another
definition:
\begin{eqnarray}
{\hat R}_i(\mu) &\equiv& \frac{1}{(M_K^{\rm  exp})^2}
 \left[m_M^2\frac{\langle{\bar K}^0|{\cal O}_i(\mu)|K^0\rangle}
 {\langle{\bar K}^0|{\cal O}_1(\mu)|K^0\rangle}\right]_{\rm
lat} ,\hspace{0.5cm} i=2,\dots,5, 
\label{eq:ri}
\end{eqnarray}
where $m_M$ is the pseudoscalar meson mass on the lattice and 
$M_K^{\rm exp}$ is the experimental kaon mass.

\begin{figure}[t]
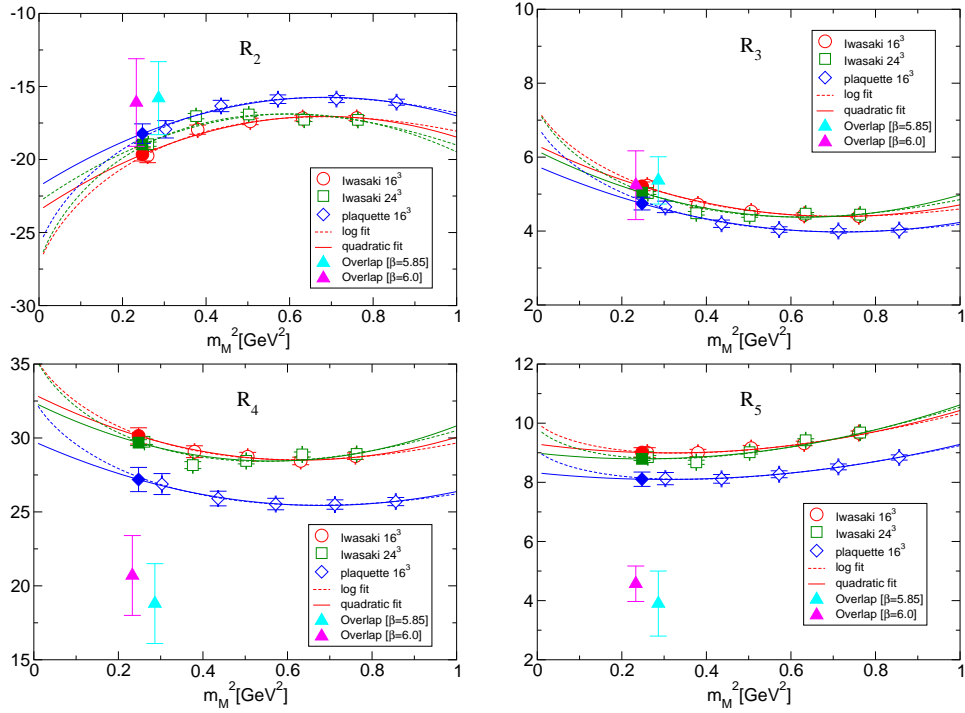

  \centering
  \includegraphics[width=60mm,clip]{fig/ch_R2_sqmass-RG.eps}
  \hspace*{5mm}
  \includegraphics[width=60mm,clip]{fig/ch_R3_sqmass-RG.eps}
  \includegraphics[width=60mm,clip]{fig/ch_R4_sqmass-RG.eps}
  \hspace*{5mm}
  \includegraphics[width=60mm,clip]{fig/ch_R5_sqmass-RG.eps}
  \caption{${\hat R}_i$ as a function of $m_M^2$. 
Solid symbols denote the extrapolated values at $M_K^{\rm exp}$.
The lattice spacing is fixed by the Sommer scale.}
  \label{fig:Ri_comp}
\end{figure}

In Figure~\ref{fig:Ri_comp} we plot the results for ${\hat R}_i$ as a function
of $m_M^2$.
We observe little volume dependence for the Iwasaki gauge action.
The results of the Iwasaki and the plaquette gauge
actions show 10\% discrepancies, which could be $O(a^2)$ effects due to
the difference of the gauge actions.
We extrapolate the results for ${\hat R}_i$ at the physical kaon mass,
which is located around $m_fa\sim0.02$,
employing the quadratic and the logarithmic fitting functions:
\begin{eqnarray}
{\hat R}_i &=& b_0+b_1m_M+b_2(m_M)^2, \\
{\hat R}_i &=& b_0+b_1m_M+b_2m_M\log{(m_M)}.
\end{eqnarray}
The results are depicted in Figure~\ref{fig:Ri_comp}, where
both fitting functions give consistent results
at the physical point.   
Table~\ref{tab:Ri_phys} summarizes 
the values of ${\hat R}_i$ extrapolated at the physical point
with the quadratic fitting function.
The first errors are statistical. The second denote differences
of the central values for the two fitting functions, 
which are considered to be the systematic errors.
We check numerically 
that the uncertainties due to a choice of the fitting functions
are smaller than the statistical errors. 
The results for ${\hat R}_i$ show a striking feature that
the non-SM matrix elements are much
larger than the SM ones for both gauge actions.
Especially, the magnitudes of the matrix elements 
$\langle{\bar K}^0|{\cal O}_2(\mu)|K^0\rangle$
and
$\langle{\bar K}^0|{\cal O}_4(\mu)|K^0\rangle$
are one order larger than that of 
$\langle{\bar K}^0|{\cal O}_1(\mu)|K^0\rangle$.

\begin{table}[t]
\begin{center}
\caption{Results for ${\hat R}_i$ and $B_i$ extrapolated at the physical point 
with the quadratic fit. The renormalization scale is $\mu=1/a$}
\label{tab:Ri_phys}
\begin{tabular}{c|l|ll}
\hline
\hline
&\multicolumn{1}{|c|}{plaquette} & \multicolumn{2}{c}{Iwasaki}  \\
$N_s\times N_t\times N_5$ & \multicolumn{1}{|c|}{$16^3\times32\times16$}   
 & \multicolumn{1}{c}{$16^3\times32\times16$} & \multicolumn{1}{c}{$24^3\times32\times16$}  \\ 
\hline
${\hat R}_1$ & \multicolumn{1}{|c|}{$1$} & \multicolumn{1}{c}{$1$}  & \multicolumn{1}{c}{$1$} \\ 
${\hat R}_2$ & $-18.22(69)(38)$ & $-19.65(45)(21)$ & $-18.97(16)(16)$   \\ 
${\hat R}_3$ & $\phantom{-}\phantom{0}4.74(17)(10)$ & $\phantom{-}\phantom{0}5.21(12)(6)$ & $\phantom{-}\phantom{0}5.039(41)(44)$  \\ 
${\hat R}_4$ & $\phantom{-}27.19(82)(24)$ & $\phantom{-}30.14(55)(11)$ & $\phantom{-}29.68(20)(13)$   \\ 
${\hat R}_5$ & $\phantom{-}\phantom{0}8.11(24)(6)$ & $\phantom{-}\phantom{0}9.01(16)(3)$ & $\phantom{-}\phantom{0}8.794(58)(34)$ \\ 
\hline
$B_1$ & $\phantom{-}\phantom{0}0.561(13)$ & $\phantom{-}\phantom{0}0.54361(65)$ & $\phantom{-}\phantom{0}0.5233(55)$ \\ 
$B_2$ & $\phantom{-}\phantom{0}0.5603(80)(9)$ & $\phantom{-}\phantom{0}0.5564(52)(10)$ & $\phantom{-}\phantom{0}0.5369(17)(3)$ \\ 
$B_3$ & $\phantom{-}\phantom{0}0.731(11)(1)$ & $\phantom{-}\phantom{0}0.7368(71)(15)$ & $\phantom{-}\phantom{0}0.7124(24)(4)$ \\ 
$B_4$ & $\phantom{-}\phantom{0}0.691(13)(8)$ & $\phantom{-}\phantom{0}0.7075(72)(31)$ & $\phantom{-}\phantom{0}0.6996(22)(26)$ \\ 
$B_5$ & $\phantom{-}\phantom{0}0.616(12)(7)$ & $\phantom{-}\phantom{0}0.6346(65)(25)$ & $\phantom{-}\phantom{0}0.6226(19)(24)$ \\ 
\hline
\hline
\end{tabular}
\end{center}
\end{table}


In Figure~\ref{fig:Bi_comp} we also show the results for the $B$-parameters 
defined in Eqs.(\ref{eq:Bk}) and (\ref{eq:Bi}).
Both gauge actions give consistent results ranging from $m_fa=0.02$ to 0.06. 
We again find little volume dependence for the Iwasaki gauge action.
In order to extrapolate $B_K$  at the physical kaon mass,
we use the following fitting function suggested by ChPT:
\begin{eqnarray}
B_K=B(1-3cm_M\log(m_M)+bm_M).
\end{eqnarray}
As for $B_i$ ($i=2,\dots,5$) we employ two fitting functions:
\begin{eqnarray}
B_i &=& b_0+b_1m_M+b_2(m_M)^2,\\
B_i &=& b_0+b_1m_M+b_2m_M\log(m_M).
\end{eqnarray}
Figure~\ref{fig:Bi_comp} 
shows that the extrapolated values at the physical point
with both fitting functions are consistent.
This is also confirmed by 
the numerical values listed in Table~\ref{tab:Ri_phys}, where
the first errors are statistical and  the second for differences
of the central values for the two fitting functions.

\section{Comparison with previous works}

In Figure~\ref{fig:Ri_comp} we compare our results for $R_i$ with those obtained
using the overlap fermion and the plaquette gauge
action\cite{Babich:2006bh}. The lattice spacing is fixed by the Sommer
scale $r_0=0.5$fm. Both results show good consistencies for $R_2$ and
$R_3$, while large deviations are observed for $R_4$and $R_5$.
 
\begin{figure}[t]
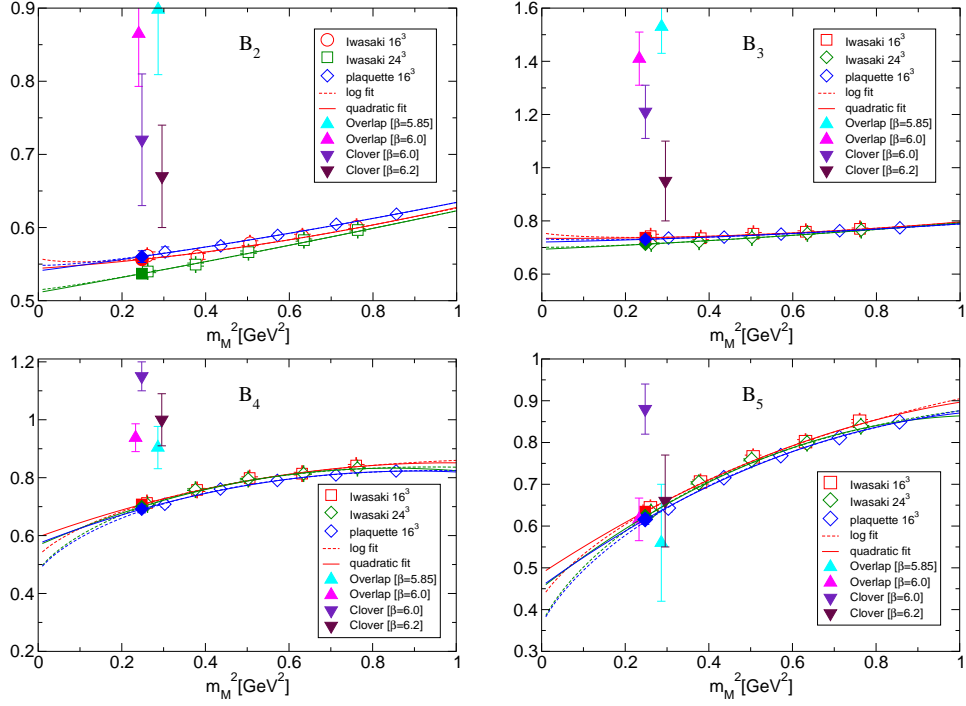

  \centering
  \includegraphics[width=60mm,clip]{fig/ch_B2-ren.eps}
  \hspace*{5mm}
  \includegraphics[width=60mm,clip]{fig/ch_B3-ren.eps}
  \includegraphics[width=60mm,clip]{fig/ch_B4-ren.eps}
  \hspace*{5mm}
  \includegraphics[width=60mm,clip]{fig/ch_B5-ren.eps}
  \caption{Bag parameters $B_i$ as a function of
$m_M^2$.Solid symbols denote the extrapolated values at
$M_K^{\rm exp}$.
The lattice spacing is fixed by the Sommer scale.}
  \label{fig:Bi_comp}
\end{figure}

As for the $B$-parameters we plot in Figure~\ref{fig:Bi_comp}
the previous results with the overlap fermion\cite{Babich:2006bh} 
and the $O(a)$-improved Wilson quark action\cite{Allton:1998sm}.
We again employ the Sommer scale to determine the lattice spacing.
Large discrepancies are observed for all the $B$-parameters except $B_5$,
though the previous results have rather large statistical errors.


Although the different choices of the quark and the gauge actions
should allow the $O(a^2)$ uncertainties, the magnitudes of 
the inconsistencies between the previous results and ours 
are more than expected. We suspect that
a possible source of discrepancies is difference of the
renormalization methods: one-loop perturbation for our results and
nonperturbative MOM scheme for the previous ones.
We are now working on a nonperturbative renormalization for the domain
wall fermion using the Schr{\"o}dinger 
functional method\cite{Taniguchi:2006qw}.


%
This work is supported in part by Grants-in-Aid for Scientific Research 
from the Ministry of Education, Culture, Sports, 
Science and Technology (Nos.~13135204,
13135216,
15540251,
16540228,
17340066, 
17540259,
17740171,
18104005, 
18540250,
18740130,
18740139).


\begin{thebibliography}{99}

\bibitem{fcnc_susy}
F.~Gabbiani, A.~Masiero,
Nucl.~Phys.~{\bf B322} (1989) 235;
J.~S.~Hagelin, S.~Kelley and T.~Tanaka,
Nucl.~Phys.~{\bf B415} (1994) 293;
F.~Gabbiani, E.~Gabrielli, A.~Masiero and L.~Silvestrini,
Nucl.~Phys.~{\bf B477} (1996) 321.


\bibitem{Babich:2006bh}
  R.~Babich {\it et al.},
  hep-lat/0605016.


\bibitem{Allton:1998sm}
  C.~R.~Allton {\it et al.},
  Phys.~Lett.~{\bf B453} (1999) 30.


\bibitem{Noaki:2001un}
  CP-PACS Collaboration, J.~Noaki {\it et al.},
  Phys.~Rev.~{\bf D68} (2003) 014501.


\bibitem{necco}
S.~Necco and R.~Sommer,
Nucl.~Phys.~{\bf B622} (2002) 328.

\bibitem{takeda}
CP-PACS Collaboration, S.~Takeda {\it et al.},
Phys.~Rev.~{\bf D70} (2004) 074510.

\bibitem{dwf_bk}
  CP-PACS Collaboration, A.~Ali~Khan {\it et al.},
  Phys.~Rev.~{\bf D64} (2001) 114506.

\bibitem{dwf_chiral}
  CP-PACS Collaboration, A.~Ali~Khan {\it et al.},
  Phys.~Rev.~{\bf D63} (2001) 114504.

\bibitem{pt_dwf}
S.~Aoki, T.~Izubuchi, Y.~Kuramashi and Y.~Taniguchi,
Phys.~Rev.~{\bf D59} (1999) 094505; {\it ibid.}~{\bf D60} (1999) 114504; 
{\it ibid.}~{\bf D67} (2003) 094502.


\bibitem{Nakamura:2006zx}
  Y.~Nakamura and Y.~Kuramashi,
  Phys.~Rev.~{\bf D73} (2006) 094502.



\bibitem{Taniguchi:2006qw}
  Y.~Taniguchi,
  hep-lat/0604002.

\end{thebibliography}
\end{document}